\documentclass[fleqn,usenatbib]{mnras}

\usepackage{newtxtext,newtxmath}

\usepackage[T1]{fontenc}
\usepackage{ae,aecompl}
\usepackage[utf8]{inputenc}
\usepackage{soul}
\usepackage{float}
\usepackage{multicol}

\usepackage[switch]{lineno}
\usepackage{ulem}
\usepackage{xcolor}

\usepackage{graphicx}	
\usepackage{amsmath}	
\usepackage{amssymb}	

\title[DeepCuts]{Classifying CMB time-ordered data through deep neural networks}

\author[F. Rojas et al.]{
Felipe Rojas,$^{1}$\thanks{E-mail: flrojas@uc.cl}
Lo\"ic Maurin$^{2}$,
Rolando D\"{u}nner$^{3}$,
and Karim Pichara$^{1,4}$
\\
$^{1}$Departamento de Ciencia de la Computaci\'on, Facultad de Ingenier\'ia, Pontificia Universidad Cat\'olica de Chile,\\
Av. Vicu\~na Mackenna 4860, 7820436, Macul, Santiago, Chile\\
$^{2}$Universit\'e Paris-Saclay, CNRS,  Institut d$'$astrophysique spatiale, 91405, Orsay, France\\
$^{3}$Instituto de Astrof\'isica and Centro de Astro-Ingenier\'ia, Facultad de F\'isica, Pontificia
Universidad Cat\'olica de Chile,\\ Av. Vicu\~na Mackenna 4860, 7820436, Macul, Santiago, Chile\\
$^{4}$ Millennium Institute of Astrophysics. Santiago, Chile
}

\date{Accepted XXX. Received YYY; in original form ZZZ}

\pubyear{2020}

\begin{document}
\label{firstpage}
\pagerange{\pageref{firstpage}--\pageref{lastpage}}
\maketitle

\begin{abstract}
The Cosmic Microwave Background (CMB) has been measured over a wide range of multipoles. Experiments with arc-minute resolution like the Atacama Cosmology Telescope (ACT) have contributed to the measurement of primary and secondary anisotropies, leading to remarkable scientific discoveries. Such findings require careful data selection in order to remove poorly-behaved detectors and unwanted contaminants. The current data classification methodology used by ACT relies on several statistical parameters that are assessed and fine-tuned by an expert. This method is highly time-consuming and band or season-specific, which makes it less scalable and efficient for future CMB experiments. In this work, we propose a supervised machine learning model to classify detectors of CMB experiments. The model corresponds to a deep convolutional neural network. We tested our method on real ACT data, using the 2008 season, 148 GHz, as training set with labels provided by the ACT data selection software. The model learns to classify time-streams starting directly from the raw data. For the season and frequency considered during the training, we find that our classifier reaches a precision of 99.8\%. For 220 and 280 GHz data, season 2008, we obtained 99.4\% and 97.5\% of precision, respectively. Finally, we performed a cross-season test over 148 GHz data from 2009 and 2010 for which our model reaches a precision of 99.8\% and 99.5\%, respectively. Our model is about 10x faster than the current pipeline, making it potentially suitable for real-time implementations.
\end{abstract}




\section{Introduction}
Over the last years, many efforts have been done in order to measure the Cosmic Microwave Background (CMB) temperature and polarization anisotropies. Telescopes like the Atacama Cosmology Telescope (ACT,~\cite{Swetz_2011}), South Pole Telescope (SPT,~\cite{Ruhl2004}), Polarbear~\citep{Lee2008}, among others, have mapped the microwave sky in temperature with unprecedented accuracy using multi-detector arrays, providing measurements of primary and secondary perturbations (e.g.~\cite{Das2011,Hand2012}). More recently, new detector arrays sensitive to linear polarization at multiple frequency bands (e.g.~\cite{Niemack2010,Austermann_2012}) are complementing the previous science through high-sensitivity observations of the CMB polarization, providing additional information to probe cosmological physics (for an overview, see~\cite{abazajian2016cmbs4}). There are big efforts in this direction. For example, the future Simons Observatory~\citep{Ade_2019,galitzki2018simons} will map the millimeter sky, both temperature and polarization, in six bands: 39, 93, 145, 225 and 280 GHz, through a set of three small apertures telescopes (0.5-m) and one large aperture telescope (6-m), targeting large and arc-minute scales, respectively. Such configuration includes 30,000 bolometers for the large aperture telescope and another 30,000 distributed in the three small aperture telescope. For CMB-S4~\citep{abitbol2017cmbs4}, the scientific goals require even more detectors (\textit{O}(100,000)) to achieve a sensitivity of order 1 $\mu$K-arcmin, generating large volumes of raw data.

From the data analysis perspective, the challenge is important. Ground-based telescopes like ACT scan the sky with arrays of thousands of detectors, generating time-streams that are stored as files of fixed length called time-ordered data (TOD). After data acquisition, the TOD files need to be selected, calibrated, and mapped. The data selection step implies that vast amount of time-streams need to be analyzed to remove poorly-behaved detectors, glitches, and other unwanted contaminants that may affect the quality of the final maps. Up to now, ACT relies on a methodology based on statistical tests that are assessed by expert software, which rejects full TOD files, individual detectors, and segments of them based on several statistical estimators~\citep{Dunner2013}. This method is intrinsically arbitrary, as it implies a decision on whether the quality of the data is good enough to be incorporated in the maps. In order to achieve a good data selection, the process is iterated with map-making to make sure that the residual contamination is at acceptable levels given the experimental goals. This is done for each band and season separately, implying that an important amount of time is spent selecting data, making the entire process less suitable for future projects. Additionally, there are some assumptions behind the current method that could not hold for future experiments. In particular, the method utilizes the atmospheric signal to determine whether the detectors are coupled to the optical signal and to compute the relative calibration between them. The correlation with the atmosphere is useful for frequencies where its brightness is high enough (150 GHz and higher) but could be uninformative for lower frequencies where the atmosphere signal weakens, and effects like the thermal drift of the cryostat start to dominate.
Then, there is the necessity of more robust and scalable algorithms to face the forthcoming challenges regarding CMB data selection.

In this work, we propose a machine learning approach to the CMB data selection problem. Specifically, we designed a deep residual neural network that takes individual detector time-streams as the input and predicts the probability of two classes of detectors: good or bad. The model automatically designs high order statistics without human intervention. The method was tested on the ACT MBAC dataset, comparing its performance to the existing expert system, and assessing its ability to operate across data from different seasons and observing frequency bands. To the best of our knowledge, this is the first effort intended to perform the classification of raw CMB data using a machine learning technique that does not require manual feature engineering.

The paper is organized as follows. Section~\ref{sec:background} gives an overview of ACT and details about the current data selection methodology. We also describe briefly deep learning and some applications, and mathematical details concerning convolutional neural networks. Section~\ref{sec:model} explains the proposed model and its design. Section~\ref{sec:data} describes the data sets used for training and testing. The results are presented in Section~\ref{sec:results}. Finally, we discuss our results in Section~\ref{sec:discussion}.

\section{Background}\label{sec:background}
In this section, we describe in details the CMB data, how current ACT data selection methodologies work, and we also provide the basics on convolutional neural networks.

\subsection{General CMB data description}\label{sec:cmb_general}
Most current CMB experiments use cameras composed of hundreds to thousands of individual bolometric detectors. They record the total power --over a broad passband-- of incoming radiation integrated over the optical beam of the telescope projected on the sky. The leading detector technology is transition/edge sensors (TES), which make use of a superconducting transition to detect slight changes in optical loading. 
Thus they require very low and stable temperatures, of a few mK, being susceptible to thermal noise contamination (see e.g.~\cite{Dunner2013}). 
The telescope scans the sky in azimuth, typically at constant elevation strategy to avoid changes in sky loading.
Each detector samples the sky power at a constant rate, producing a time-stream recorded in sync with the pointing information and other relevant housekeeping data.
The final CMB maps are then solved from all the samples in a season, their pointing information and a model of the noise and systematics.
For storage and analysis convenience the time-streams are usually stored in chunks, with lengths of a few minutes, which is the time over which the observing conditions can be considered stationary. Information on longer time scales is not relevant for the analysis as the telescope operation is not stable.

For our study we used data from the Atacama Cosmology Telescope (ACT), obtained between the years 2008 and 2010 by the 3-in-1 Millimiter Bolometer Array Camera (MBAC). MBAC operated in three frequency bands, centered at 148, 218 and 277 GHz, with 1024 detectors per band, and was polarization insensitive. The detectors are distributed in arrays of 32x32 pop-up TES bolometers. Each of those are read out through Superconducting Quantum Interference Device (SQUID) multiplexers. The bolometers signal is amplified by an array of SQUIDs controlled by Multi-Channel Electronics (MCEs,~\cite{2008SPIE.7020E..28B}) independently for each band. The cryogenic system comprises a first stage with two pulse-tube coolers, used to cool down the optics. Then, there are two $^4$He sorption fridges: the first one designed to cool down all 1K optical components, while the second is used to precool and back the final stage refrigerator. Finally, there is a $^3$He sorption fridge to reach the base temperature of 300 mK. For further details, see~\cite{Swetz_2011}.

\begin{figure}
    \centering
	\includegraphics[width=\columnwidth,keepaspectratio=true]{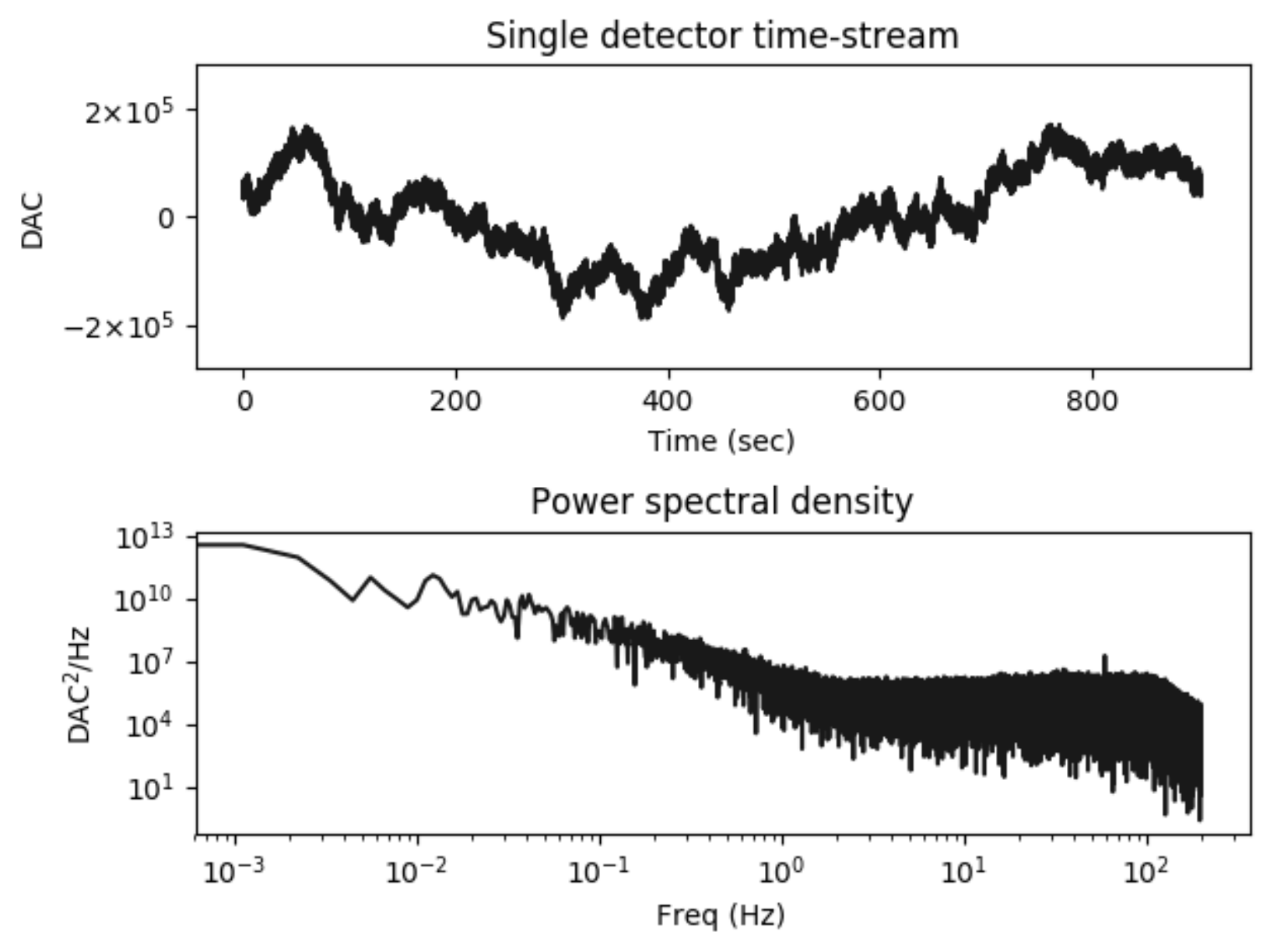}
    \caption{Single detector time-stream (upper panel) and its power spectral density (lower panel).}
    \label{fig:detpsd}
\end{figure}

At the wavelengths of interest, the signal is dominated by the atmospheric emission, mostly from precipitable water vapor (PWV), with a Rayleigh-Jeans temperature of a few Kelvin.
This signal is modeled by atmospheric turbulence, following a Kolmogorov spectrum~\citep{Tatarski1961,Church_1995,Lay_2000,Dunner2013} which falls as a power-law of index -3.7 in spatial frequency.
Modulated by the telescope scan, the atmosphere signal builds up at low TOD frequencies.
In contrast, the CMB signal is very faint, of order a hundred micro-Kelvin or less, being modeled by the CMB power spectrum, with relevant TOD frequencies of up to a few tens of Hertz.
This signal is buried under the detector's thermal noise, which for MBAC ranged between 1 and 2 milli-Kelvin, and cryogenic thermal fluctuations described by a $1/f$ power law.
Fig~\ref{fig:detpsd} shows the typical signal seen by a MBAC detector. The low frequency part is dominated by atmospheric emission and slow drifts of the thermal bath of the detectors. It can be described by a power law with a varying spectral index depending on the changing properties of the atmosphere. The high frequency part is dominated by detector noise and is well described by a white noise plateau.

\subsection{Current data selection methodology}\label{sec:cuts}
In a regime of high number of detectors ($\sim 10^3$) and large integration time (several months to several years), we need to select the data that will be used to map the sky. On average, about $30\%$ of the data is not usable, mostly because of bad weather condition or because individual detectors are not operating as they should. The current methodology to reject data that should not be mapped relies on our understanding of the instrument, its environment and the signal from the sky. Expert systems are built to statistically analyze and reject data. 

For ACT we use the characteristics of the data described above: at low-frequencies the data is dominated by the atmosphere signal and $1/f$ noise, while at high-frequencies it is dominated by detector noise. Using these assumptions, the rejection pipeline essentially evaluates if a given detector is seeing the atmosphere and if its noise properties are consistent with our model in terms of sensitivity and Gaussianity. 
Around ten statistical estimators are computed and ranges of acceptable values are defined for each of them based on the value distributions for a whole year of data.
The estimators are designed and tuned taking into consideration all the available knowledge of the system.
A key test to check if a detector is coupled to the optical signal is to determine if it sees the atmosphere.
Since we lack of an independent measurement of the atmospheric signal, it needs to be estimated from the data itself.
Considering that the Kolmogorov spectrum is dominated by spatially large modes, the atmosphere is expected to produce a strong common mode across the focal plane of each detector array, which are a third of a degree across for MBAC, in sync with the telescope scan.
This low-frequency common mode is then used as a template for the atmosphere signal which can be correlated to independent detectors to determine if they are coupled to the incoming radiation.
This method is limited by the strong $1/f$ thermal contamination at low-frequencies, which is mitigated by subtracting the thermal common mode measured by dark detectors not coupled to radiation.
Other important estimators are used to determine if individual detectors have the expected high-frequency noise level and if it is Gaussian.
A more complete description can be found in~\cite{Dunner2013}.

The expert system has been used successfully for ACT data analysis and public releases. Nevertheless, it relies strongly on human intervention for fine-tuning of the model parameters, systematic effects mitigation and setting thresholds for acceptable values of the statistical estimators, which adds uncertainty to the labels. Then, data selection results can be unstable with respect to the choices made in the design of the system. To minimize the uncertainty over the labels, the system is fine-tuned by iterating between the data selection and the maps, which are assessed in terms of the final noise gaussianity and systematic effects. This is highly time-consuming and difficult to scale across seasons and detector arrays. Our proposal is to accelerate this process by learning a mapping function between the data and the labels to perform automatic classification. Although there is an intrinsic bias due to the noisy expert labels, an automatic classifier could boost the data quality assessment and enable real-time failure detection.

\subsection{Deep Learning}\label{sec:dl}
Deep learning has gathered much attention due to its ability to tackle difficult tasks like natural images classification~\citep{He2016}, speech recognition~\citep{Chiu2018}, natural language processing~\citep{young2017recent}, data generation~\citep{kingma2013autoencoding}, and more. Its power relies on a stack of several non-linear modules that can model complex structures in the data and design useful features automatically~\citep{LeCun2015a}.

\par
The use of this technology in astronomy is extensive. The increasing complexity and size of astronomical databases require the development of scalable pipelines, reducing human intervention and improving efficiency. Some recent advances include variable star classification~\citep{Aguirre_2018}, gravitational wave detection~\citep{GEORGE201864}, image denoising~\citep{Schawinski_2017}, stellar parameters estimation~\citep{Yang2015}, exoplanet identification~\citep{Shallue2018}, and many more. In the CMB field, the current developments are focused mainly on the final maps. For example, the work of~\cite{PERRAUDIN2019130} proposed a deep learning model called DeepSphere that is capable of predicting a class from a map, predict parameters from a map, classify pixels, and predict a set of maps from a map. Similarly,~\cite{Krachmalnicoff_2019} presents a convolutional neural network on the HEALPix sphere, applying it to the prediction of cosmological parameters. In~\cite{Caldeira_2019}, a UNet-like architecture is proposed to tackle the reconstruction of the CMB lensing potential. The work of~\cite{mnchmeyer2019fast} proposes the WienerNet, a neural network that learns to filter masked CMB maps, being about 1000 times faster than current methods based on the conjugate gradient. Although the literature shows that deep learning methods are being adopted and used on CMB products like the maps, there is a lack of research regarding automatic classification of CMB time-streams. In this work we aim to fill this gap.

\subsubsection{Convolutional Neural Networks}

Many modern deep learning architectures are based on convolutions, one of the most used building blocks. Convolutional layers are widely used in images, but they can be applied to time series as well. In this type of network, each layer contains a filter bank that is connected to the features produced by the previous layer. The main advantage of these layers is the fact that the filters have a limited field of view, which means that their connections are local and share weights, decreasing the complexity in comparison to the traditional neural networks based on matrix multiplications. The local connections allow to learn specific patterns that can repeat across the data. Below, we describe the traditional formulation of the convolution operation in the context of neural networks. Then, we present a variation that is called depth-wise separable convolution, which is the definition used in this work.

Consider a time series $\mathbf{x}^{l}_{i}(t) = \lbrace \mathbf{x}_{i1}, \mathbf{x}_{i2}, ... , \mathbf{x}_{in}\rbrace$ where $i$ is the index of the channels with $i = 1, ..., n_{ch}$, and $n$ is the length; a set of filters $w_j^l$ of fixed size $f_s$ with $j = 1, ..., n_{f}$ and $l = 1, ..., L$, the number of filters and layers, respectively. For a given layer $l$, the input $\mathbf{x}^{l-1}_{i}$ is convolved with the filters according to
\begin{equation}\label{eq:conv}
\mathbf{h}^l_j(t) = \sigma\left(\sum_{i} w_{j}^{l} * \mathbf{x}^{l-1}_{i}(t) + b_{j} \right)
\end{equation}
 
In this formulation, the output of each filter $j$ is computed as the sum of each input channel convolved with that filter plus a bias term $b_j$. The free parameters here are $w_j^l$ and $b_j$. Notice that the kernel $w_j^l$ has the shape $(f_s,i,j)$. The function $\sigma$ is an activation function that is set at the beginning. Typically, deep learning models use activations like the rectified linear units (ReLU,~\cite{alexnet}) and its variants like Parametric ReLU~\citep{He2015} and Leaky ReLU~\citep{Maas13rectifiernonlinearities}. The role of this activation function is to force the algorithm to find non-linear and non-trivial patterns and relationships in the data that traditional methods are not able to model.

There is a variation of the definition~\ref{eq:conv} in which each input channel is convolved separately with a single kernel in a first step called depth-wise convolution and then, the result is combined in a second step called point-wise convolution, where the kernel operates along the channel axis with independent filters of length equal to the number of input channels; the number of filters defines the amount of output channels. In other words, there are two kernels to be learned: the first, corresponding to the depth-wise step, with shape $(f_s,i,1)$; the second, for the point-wise step, with shape $(1,i,j)$. This decomposition let us reduce the amount of free parameters per layer and accelerate the convergence during the training stage, as has been demonstrated in several works (see e.g.~\cite{Chollet2017} and references therein). This variation of the convolution operation is adopted in this work.

\begin{figure}
    \centering
	\includegraphics[width=0.8\columnwidth,keepaspectratio=true]{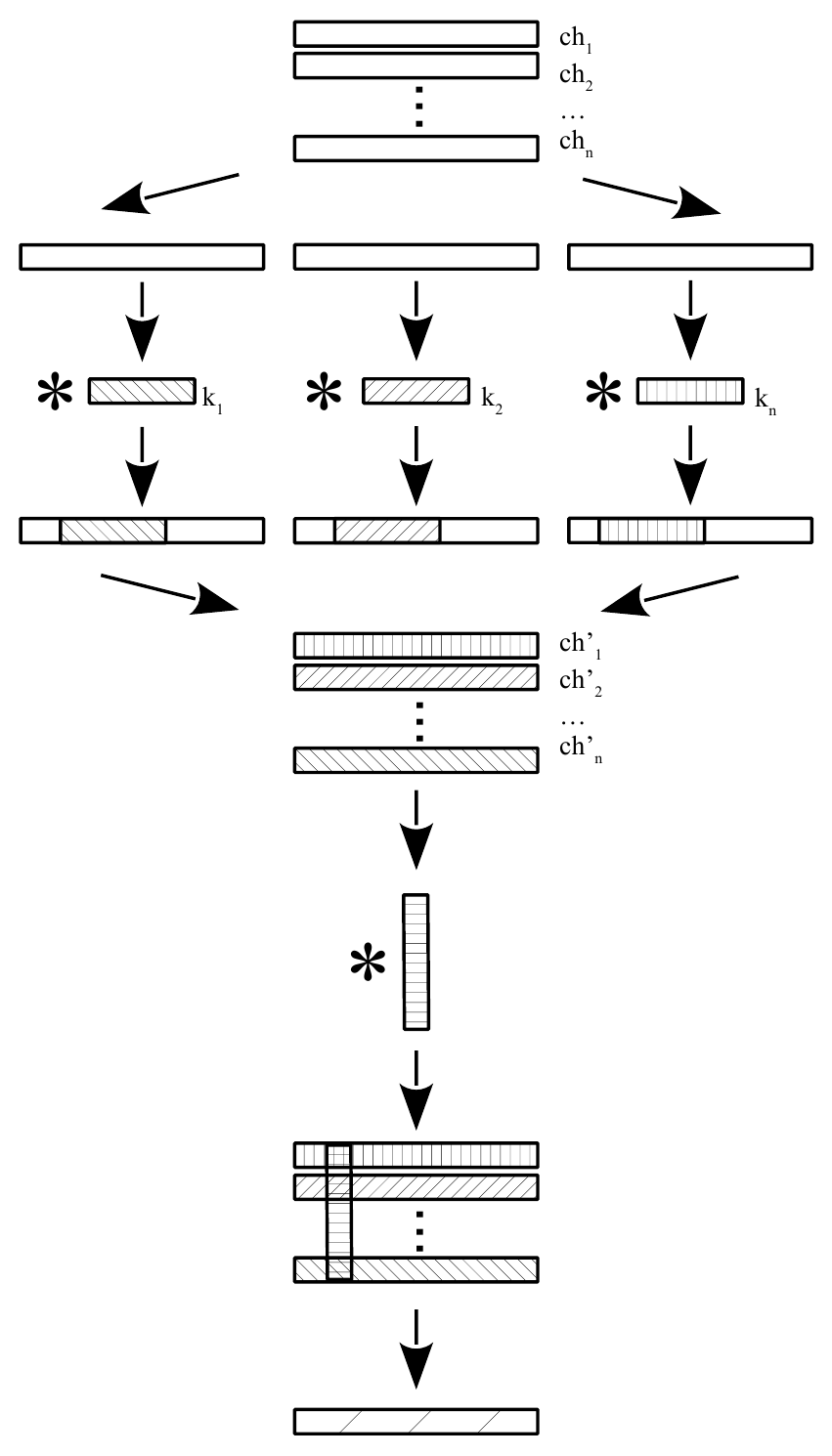}
    \caption{Separable Convolution diagram. An input with $n$ channels is split so that each channel is convolved separately. The resulting features are then convolved point-wise.}
    \label{fig:sepconv}
\end{figure}

In order to learn the best parameters for our neural network, it is necessary to define a cost function to be optimized during the training phase. This cost function tell us how well our model is estimating the relation between the input (time-stream) and the output (label), by computing the distance between the predicted label and the actual label. For classification tasks, the usual cost functions are the binary and the categorical cross entropy losses, depending on the number of classes. The cross entropy is defined as follows
\begin{equation}\label{eq:crossentr}
    H = -\sum_{i}^{n_c} t_i\log f(s_i)
\end{equation}
where $t_i$ and $s_i$ are the ground-truth label and the predicted score of class $i$, respectively; $n_c$ is the number of classes, and $f$ is an activation function that maps the high-level features to the range $[0,1]$. For the binary classification problem ($n_c=2$), the function $f$ often used is the sigmoid function: $f(s_i) = (1+\exp({-s_i}))^{-1}$, which leads to a cost function given by
\begin{equation}
    H = -t_1\log f(s_1) - (1-t_1)\log (1-f(s_1))
\end{equation}
For multi-class problems, the preferred activation function $f(s_i)$ in equation~\ref{eq:crossentr} is the softmax function: $f(s_i) = e^{s_i}(\sum_j^{n_c}e^{s_j})^{-1}$.

\section{Proposed model}\label{sec:model}
Our model consists of a stack of blocks which are composed by two separable convolution layers and a residual connection. We chose the separable convolution for two main reasons. In first place, since this type of convolution is a factorized version of the normal convolution (Section~\ref{sec:dl}), it allows us to reduce the number of free parameters per layer, reducing the complexity of our model. Secondly, we found empirically that, for our specific problem, separable convolutions outperform the classification accuracy in comparison to normal convolutions. 

The separable convolution residual block can be described by the following operations
\begin{eqnarray}
\mathbf{h}_1^{(l)} &=& \sigma\left( \mathbf{w}_1^{(l)} * \mathbf{x}^{(l-1)} + \mathbf{b}^{(l)}_1 \right)\\
\mathbf{h}_2^{(l)} &=& \sigma\left( \mathbf{w}_2^{(l)} * \mathbf{h}_1^{(l)} + \mathbf{b}^{(l)}_2 \right)\\
\mathbf{h}_3^{(l)} &=& \mathbf{x}^{(l-1)}-\mathbf{h}_2^{(l)} \\
\mathbf{h}_4^{(l)} &=& \sigma\left( \mathbf{h}_3^{(l)} \right) \\
\hat{\mathbf{h}}^{(l)} &=& \text{MaxPooling}\left( \mathbf{h}_4^{(l)} \right)
\end{eqnarray}

We have omitted the summation symbol to simplify the notation. Then, the model is written as follows
\begin{eqnarray}
\hat{\mathbf{h}}^{(0)} &=& \sigma\left( \mathbf{w}^{(0)} * \mathbf{x} + \mathbf{b}^{(0)} \right)\\
\hat{\mathbf{h}}^{(1)} &=& \text{ResidualBlock}^{(1)}\left( \hat{\mathbf{h}}^{(0)} \right)\\
\hat{\mathbf{h}}^{(2)} &=& \text{ResidualBlock}^{(2)}\left( \hat{\mathbf{h}}^{(1)} \right)\\
&\vdots& \nonumber \\
\hat{\mathbf{h}}^{(n)} &=& \text{ResidualBlock}^{(n)}\left( \hat{\mathbf{h}}^{(n-1)} \right)\\
\hat{\mathbf{h}}_{gp} &=& \text{GlobalMaxPooling}\left( \hat{\mathbf{h}}^{(n)} \right)\\
\hat{\mathbf{y}} &=& \text{Softmax}\left( \hat{\mathbf{h}}_{gp} \right)
\end{eqnarray}

\begin{figure*}
	\includegraphics[width=\textwidth,keepaspectratio=true]{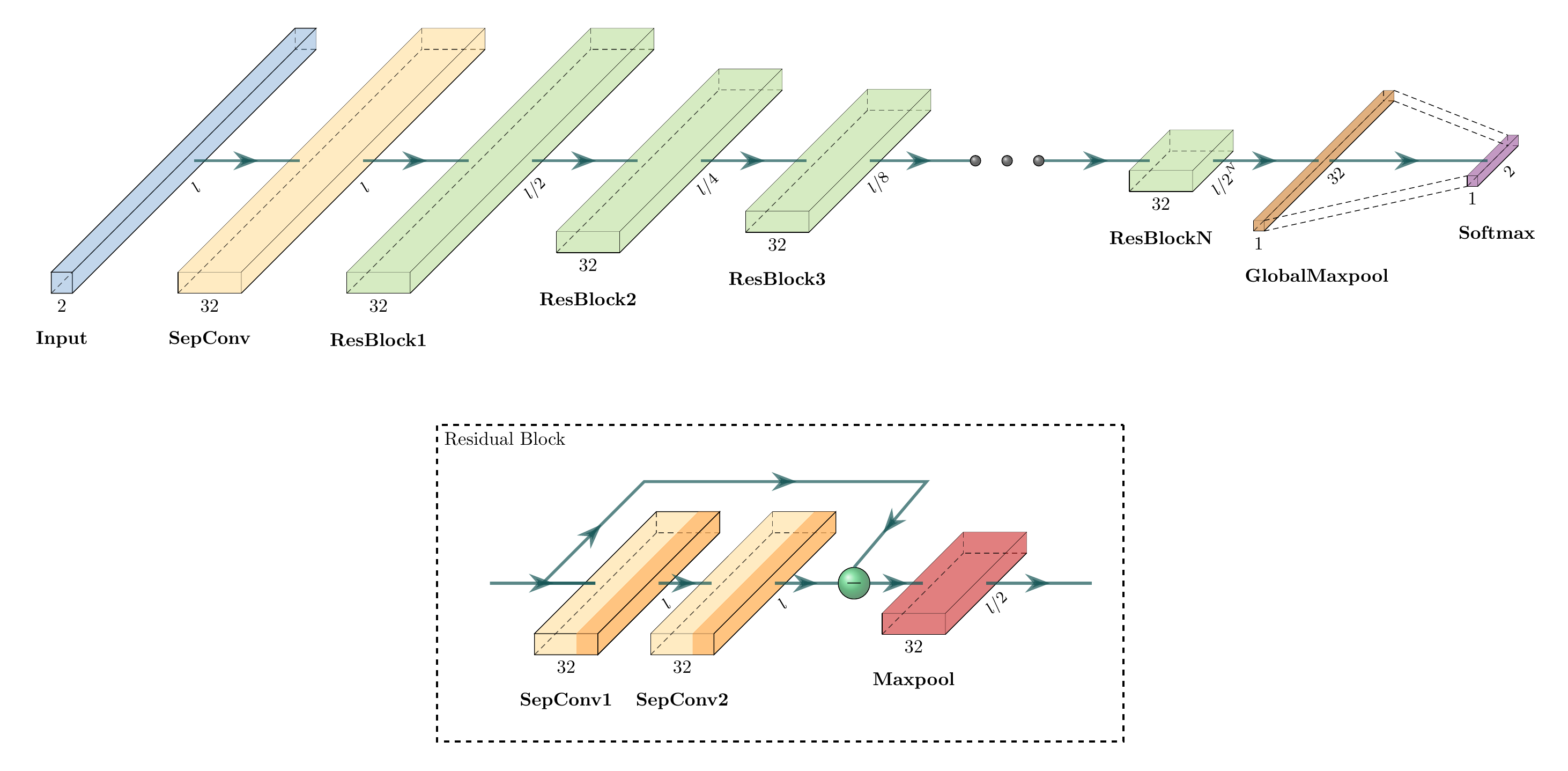}
    \caption{Model architecture. Each green block in the upper panel corresponds to a residual block, detailed in the lower panel. The darker area in the separable convolution blocks represents the activation function, which is set to linear in our work.}
    \label{fig:arch}
\end{figure*}

Figure~\ref{fig:arch} depicts the architecture of our residual network. Each residual block is composed by two separable convolution layers, one residual connection, an activation function $\sigma$, and a downsampling operator. The downsampling operation in each block is a max-pooling filter that takes the maximum of non-overlapping regions of the original input; this pooling strategy ensures that the features are invariant to translations. Notice that the first layer is not a residual block. Instead, we use a single separable convolution layer to handle the number of input channels correctly. This is done to avoid the mismatch between the number of channels in the input vector and the output of the convolutions before the residual connection. In general, the number of feature maps of each convolutional layer is greater than the number of channels of the input, requiring the mentioned expansion to allow the subtraction (or addition) in the residual step. The global pooling layer takes the maximum of each feature of the final residual block~\citep{oquab:hal-01015140}. Then, those features are reduced to two components (classes) through a dense connection which performs a matrix-vector product. Finally, the two numbers are mapped to the ground-truth by applying a softmax function. Table~\ref{tab:netconfig} summarizes the configuration of each layer.

\begin{table}
    \caption{Layers configuration. Each residual block (ResBlock) is conformed by 2 convolutional layers, an activation function, and a dowsampling step (see text). Convolutions in the same ResBlock use the same number of filters and sizes.}
    \label{tab:netconfig}
    \begin{tabular}{lcccl}
    \hline
    Layer & \# filters & Filter size & Stride & Output \\
    \hline
    Conv0 & 32 & 96 & 1 & (65536,32) \\
    ResBlock1 & 2$\times$32 & 48 & 1 & (32768,32) \\
    ResBlock2 & 2$\times$32 & 48 & 1 & (16384,32) \\
    ResBlock3 & 2$\times$32 & 48 & 1 & (8192,32) \\
    ResBlock4 & 2$\times$32 & 48 & 1 & (4096,32) \\
    ResBlock5 & 2$\times$32 & 48 & 1 & (2048,32) \\
    ResBlock6 & 2$\times$32 & 48 & 1 & (1024,32) \\
    ResBlock7 & 2$\times$32 & 48 & 1 & (512,32) \\
    ResBlock8 & 2$\times$32 & 48 & 1 & (256,32) \\
    ResBlock9 & 2$\times$32 & 48 & 1 & (128,32) \\
    ResBlock10 & 2$\times$32 & 32 & 1 & (64,32) \\
    ResBlock11 & 2$\times$32 & 32 & 1 & (32,32) \\
    ResBlock12 & 2$\times$32 & 32 & 1 & (16,32) \\
    ResBlock14 & 2$\times$32 & 16 & 1 & (8,32) \\
    GlobalPooling & - & - & - & 32 \\
    Dense & - & - & - & 2 \\
    \hline
    \end{tabular}
\end{table}

\subsection{Design details explanation}
The input vector consists of two channels. The first one, corresponds to the raw signal measured by a detector, while the second is the median value of the whole array. This median signal represents the main mode of the data for that particular observation and acts as an atmosphere template. Since our data is dominated by the atmosphere, it can be assumed in principle that all detectors correlate with it, and comparing them against a template give us information about their similarity. At the frequency bands considered in this work, good detectors look similar between them and correlate well with atmosphere, while malfunctioning ones can exhibit a drift. Then, the input of our model is designed to account for correlations between the raw detector signal and the main mode.

The residual blocks are incorporated in our model as a way to account for the deviation of the detector signal from the main mode, similar to the drift test explained in~\cite{Dunner2013}. In our experiments, we observed that the residual block is a fundamental operation for the success of our model; architectures without residual connections failed to reproduce the expert labels.

\subsection{Hyperparameters}
For the training, some parameters need to be set beforehand. In particular, we use $L_2$ regularization in the convolutional layers with coefficient 0.1. This regularization helps to avoid large values in the parameters and to keep over-fitting under control. For the optimizer, we made use of a modified stochastic gradient descent (SGD)~\citep{loshchilov2017decoupled} that shows faster convergence than the traditional SGD. It is initialized with learning rate $l_r = 0.01$ and momentum $\mu = 0.9$. The filters are initialized following a variance scaling strategy, where the values are sampled from a normal distribution with standard deviation equal to $\sigma = \sqrt{1/n}$ with $n$ the number of input units.

\subsection{Implementation}
The model was implemented using the deep learning library Keras v2.2.4~\citep{chollet2015keras} on top of Tensorflow v1.12~\citep{tensorflow2015-whitepaper}.

\section{Data sets}\label{sec:data}
The proposed model is trained using data taken by ACT during season 2008 with its 148 GHz array. The observations correspond to the southern strip defined by the limits [$20^h43^m$,$7^h53^m$] in RA and [-57.15,-48.1] in Dec., with a total area of 850 deg$^2$~\citep{Dunner2013}. The detector labels were obtained through the data selection pipeline described in Section~\ref{sec:cuts}. 

\subsection{Training, validation and testing sets}
\begin{figure}
	\includegraphics[width=\columnwidth,keepaspectratio=true]{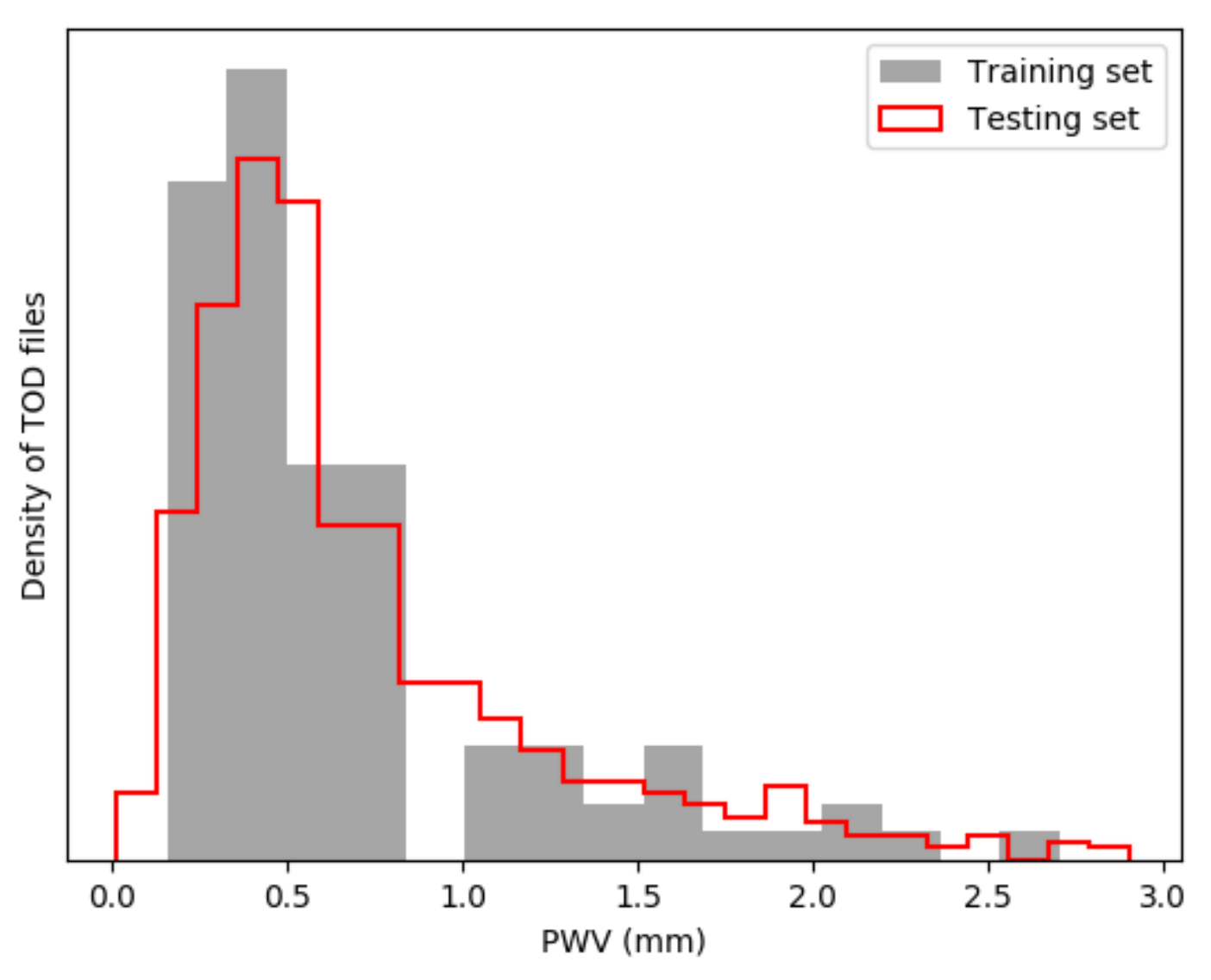}
    \caption{PWV distribution for the training and testing sets.}
    \label{fig:pwv}
\end{figure}

In order to consider a broad range of atmospheric conditions, the training and testing sets were made by selecting observations covering from August 2008 to December 2008. This ensures that the model is trained with different values of Precipitable Water Vapor (PWV), improving its capacity to generalize to different conditions. Figure~\ref{fig:pwv} shows the density of TOD files as a function of the PWV for both the training and testing sets.

The training set comprises around 114 TOD files, which represents roughly 3\% of the total amount of observations made during the mentioned season. In terms of number of detectors, these files equal to approximately 103,000 detector time-streams. We reduce this number to 47,444 by setting the maximum amount of detectors per class per TOD file as the minimum between the number of the two classes (good and bad), according to the labels provided by the expert software. This procedure help us to balance the training set (i.e., approximately equal number of samples per class) and to keep the memory and GPU requirements under control. This data set is then randomly split in proportion 80/20 for the training and validation, respectively.

For testing purposes, we have selected 637 TOD files of the same season and over the same period of time, which equals to 574,132 detector time-streams. We have also considered data from other bands and seasons to evaluate our model. These data correspond to 220 and 280 GHz, season 2008, and 148 GHz from seasons 2009 and 2010. The Table~\ref{tab:data} summarizes the data sets.

\begin{table}
    \caption{Number of detector time-streams for training, validation and testing.}
    \label{tab:data}
    \begin{tabular}{ll}
    \hline
    Frequency (Season) & Number of time-streams \\
    \hline
    148 GHz (2008) & Training: 37,955 \\
                   & Validation: 9,489 \\
                   & Testing: 574,132 \\
    \hline
    \multicolumn{2}{c}{\textit{Cross-frequency \& Cross-season tests}}\\
    \hline
    220 GHz (2008) & 225,762 \\
    280 GHz (2008) & 56,644 \\
    148 GHz (2009) & 591,807 \\
    148 GHz (2010) & 611,361 \\
    \hline
    \end{tabular}
\end{table}

\subsection{Preprocessing}\label{sec:preprocessing}
Before the training phase, the data is preprocessed according to the following steps. Each TOD file is treated separately.
\begin{enumerate}
    \item Apply IV curve calibration to transform digital units to pW.
    \item Remove dark modes. This is done to remove thermal (and/or electromagnetic) contamination. The modes considered correspond to the main mode or dark common mode, and 12 modes computed by singular value decomposition.
    \item Select a number of detectors per class equals to the minimum between the number of both classes in that file. This is done to keep the classes balanced.
    \item For each selected detector, a window of length $2^{16}$ is selected at a random position in the time-stream. The length is equivalent to 2.7 minutes of data, out of 15 minutes in total per TOD file.
    \item Each chunk is standardized by removing the mean and scaling to unit variance, detrended and tapered.
    \item The median value of all detectors of a TOD is computed (common mode). Each detector chunk is concatenated with common mode of that position.
\end{enumerate}

The atmosphere dominates the low frequency regime of our data, as mentioned in Section~\ref{sec:cmb_general}. For this reason, we concatenate each chunk with a common mode forming an input array with dimensions ($2^{16}$,2), where first dimension is the total length of the detector chunk, while the second dimension is the number of channels. We are assuming that good detectors should look similar, and their similarity can be quantified as how close to the common mode they are.

\section{Results}\label{sec:results}
In this section, we present the results on several test sets. The first classification test is done over data from band 148 GHz, season 2008, not included in the training set. Then, we report the performance of the model on different frequencies and bands.

\subsection{148 GHz, Season 2008}
The Table~\ref{tab:class_summary} summarizes the results for 148 GHz, season 2008. Notice that we have performed two tests. First, we classified the time-streams by taking the first 65,536 samples of each detector. In this case, the rate of false positives is 0.9\%, while the false negatives are 1.5\%. To evaluate whether our predictions are independent of the window, we ran a second test where we took windows at random positions. The performance is similar in both cases. 

\begin{figure*}
\begin{tabular}{cc}
    \includegraphics[width=\columnwidth,keepaspectratio=true]{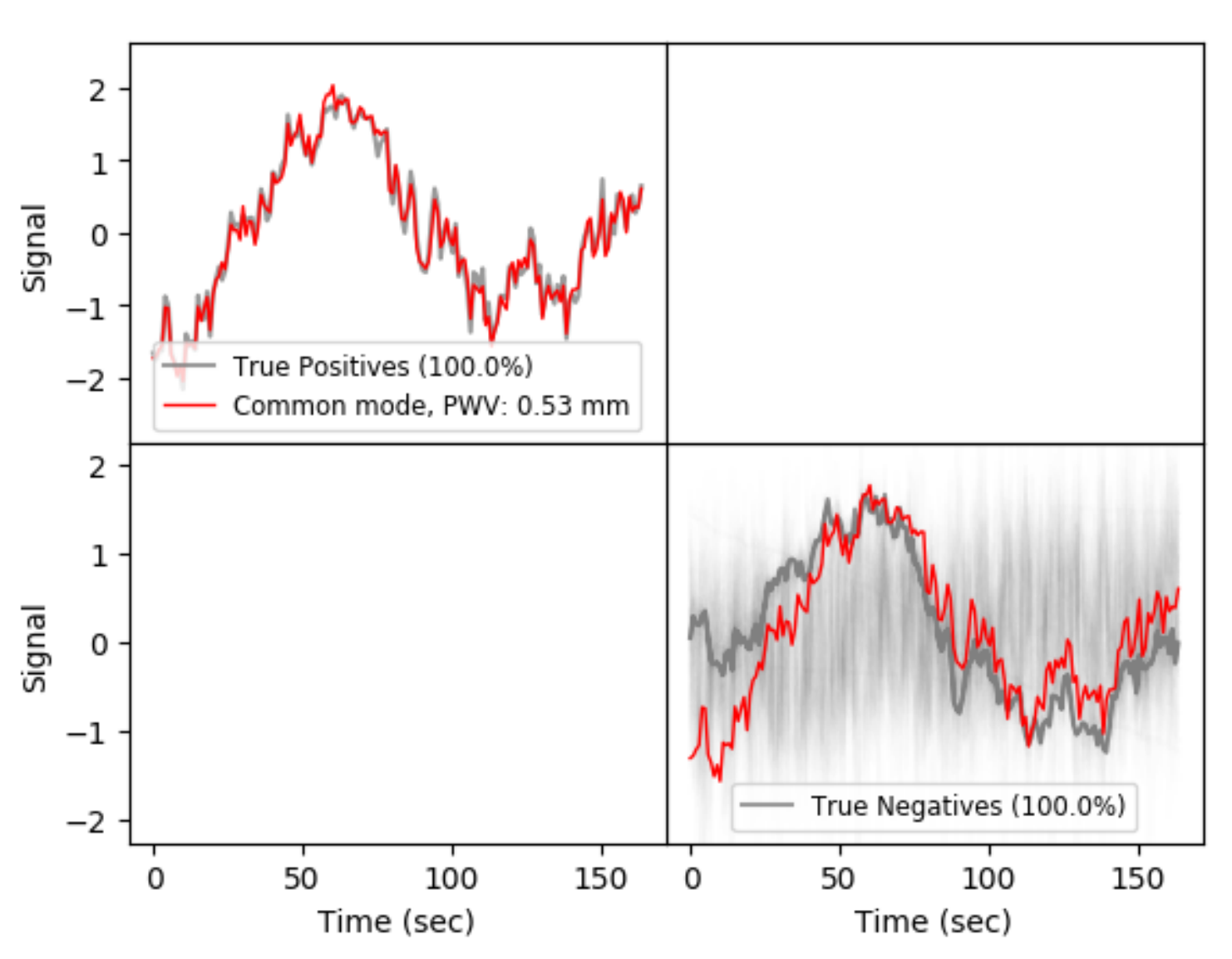} &
	\includegraphics[width=\columnwidth,keepaspectratio=true]{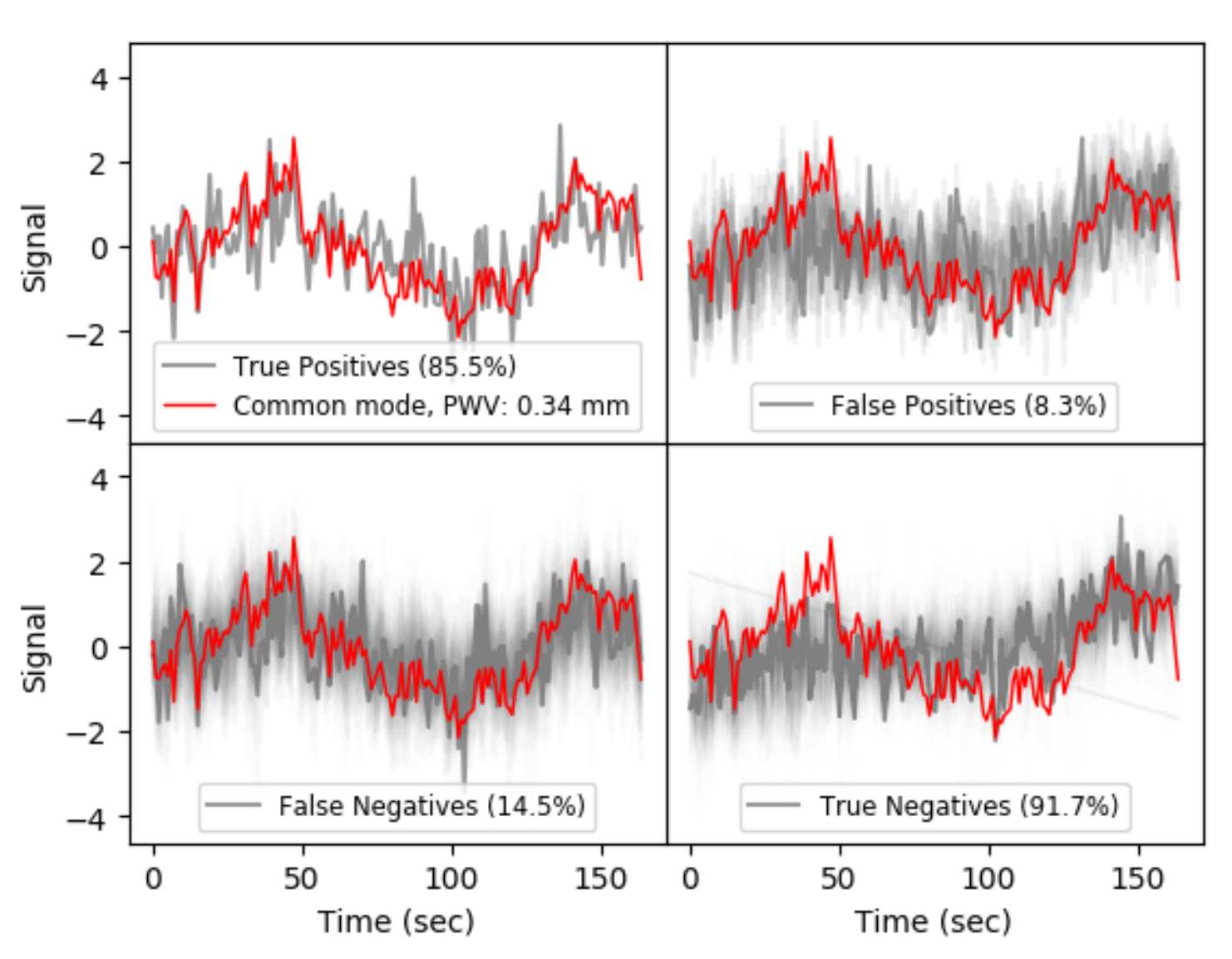} \\
\end{tabular}
    \caption{Confusion matrices showing the classification result for two TOD files. The vertical axes correspond to standardized signal, according to the preprocessing steps described in Section~\ref{sec:preprocessing}. The diffuse lines correspond to the detector time-streams according to their predicted classes. On the left panel, an observation with PWV=0.54 mm; our model was able to predict the expert labels without confusion. Notice that the true negatives do not follow the common mode. On the right panel (PWV=0.34 mm), the false positives and negatives in particular, show how difficult is to discriminate good and bad detectors for ideal observational conditions.}
    \label{fig:cmvis}
\end{figure*}

It can be shown that the error rates correlate with the PWV conditions at the time the data was taken, as lower water vapor generally imply weaker atmospheric signal available for data selection, increasing the confusion between the classes.
Figure~\ref{fig:cmvis} depicts the confusion matrices for two TOD files with PWV of 0.53 and 0.34 mm. The red line represents the common mode while the gray lines are the detector time-streams. The off-diagonal panels show the miss-classified detectors; the diagonal show the true positives and true negatives. In the case of PWV=0.53 mm, the predicted classes perfectly match the expert, in opposition to the case with PWV=0.34 mm where the model is unable to separate the classes according to the expert labels. 
On the other hand, it is known that the performance of expert system is also reduced under good PWV conditions for the same reasons, meaning that the reference labels should be considered noisier and less truth worthy.
So this result can be interpreted in three ways: either the model is doing a poor job classifying detectors, the model is doing a good job but the reference labels are noisy, or both.

\begin{table}
    \caption{Classification results on the test set for the 148 GHz band, from season 2008. The columns show the rate and number of True Positives (TP), True Negatives (TN), False Positives (FP) and False Negatives (FN).}
    \label{tab:class_summary}
    \begin{tabular}{lcccc}
    \hline
    Frequency (Season) & TP & TN & FP & FN \\
    \hline
    \multicolumn{5}{c}{\textit{First $2^{16}$ samples}}\\
    \hline
    148 GHz (2008) & 98.5\% & 99.1\% & 0.9\% & 1.5\%  \\
                   & (444,800) & (121,242) & (1,113) & (6,977) \\
                   & & & & \\
                   & Precision & 99.8\% & & \\
                   & Recall & 98.5\% & & \\
                   & F1-Score & 99.1\% & & \\
    \hline
    \multicolumn{5}{c}{\textit{Random windows of $2^{16}$ samples}} \\
    \hline
    148 GHz (2008) & 98.9\% & 98.9\% & 1.1\% & 1.1\%  \\
                   & (446,805) & (121,052) & (1,303) & (4,972) \\
                   & & & & \\
                   & Precision & 99.7\% & & \\
                   & Recall & 98.9\% & & \\
                   & F1-Score & 99.3\% & & \\
    \hline
    \end{tabular}
\end{table}

\begin{figure*}
\begin{tabular}{cc}
    \includegraphics[width=\columnwidth,keepaspectratio=true]{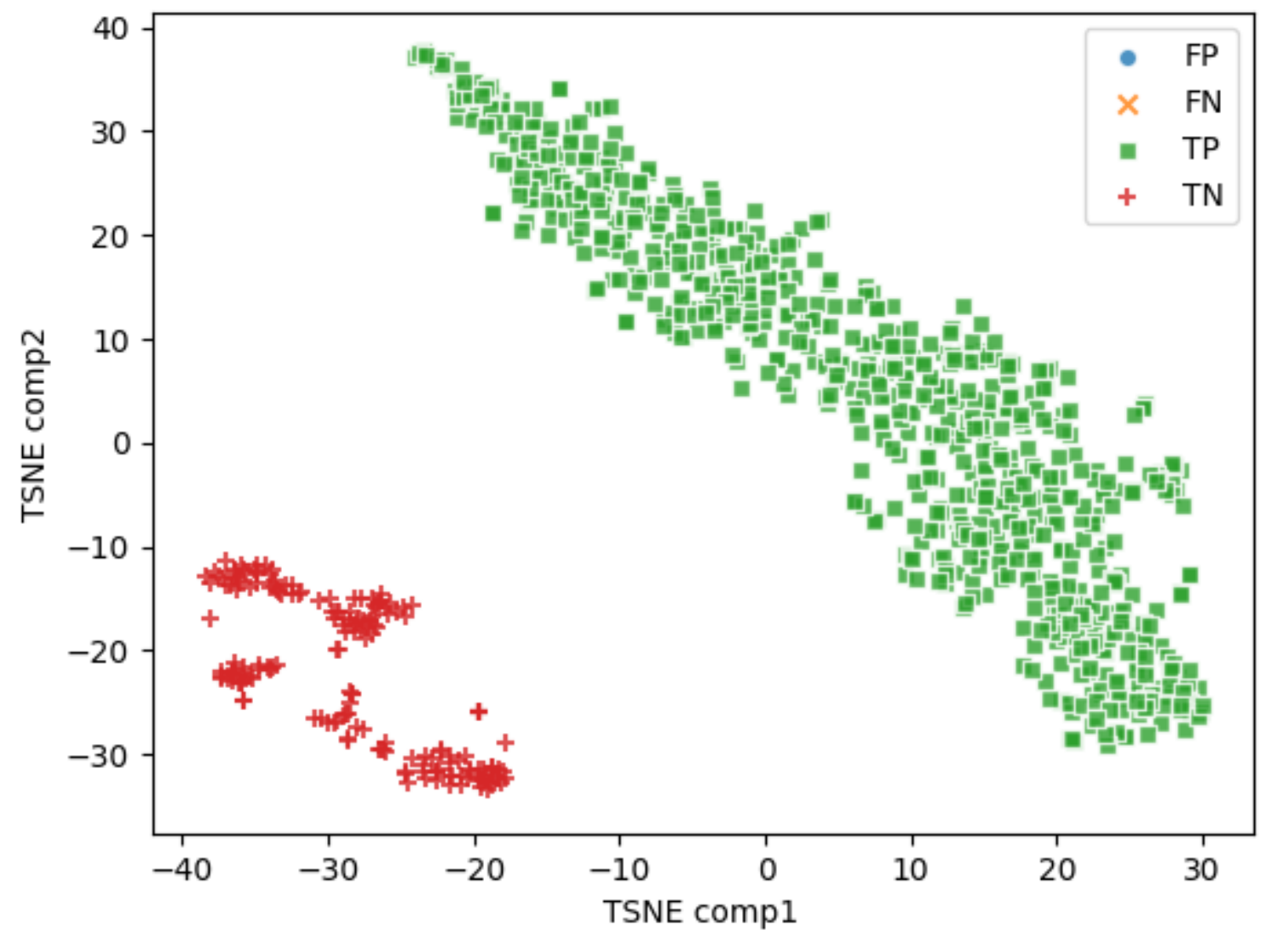} &
	\includegraphics[width=\columnwidth,keepaspectratio=true]{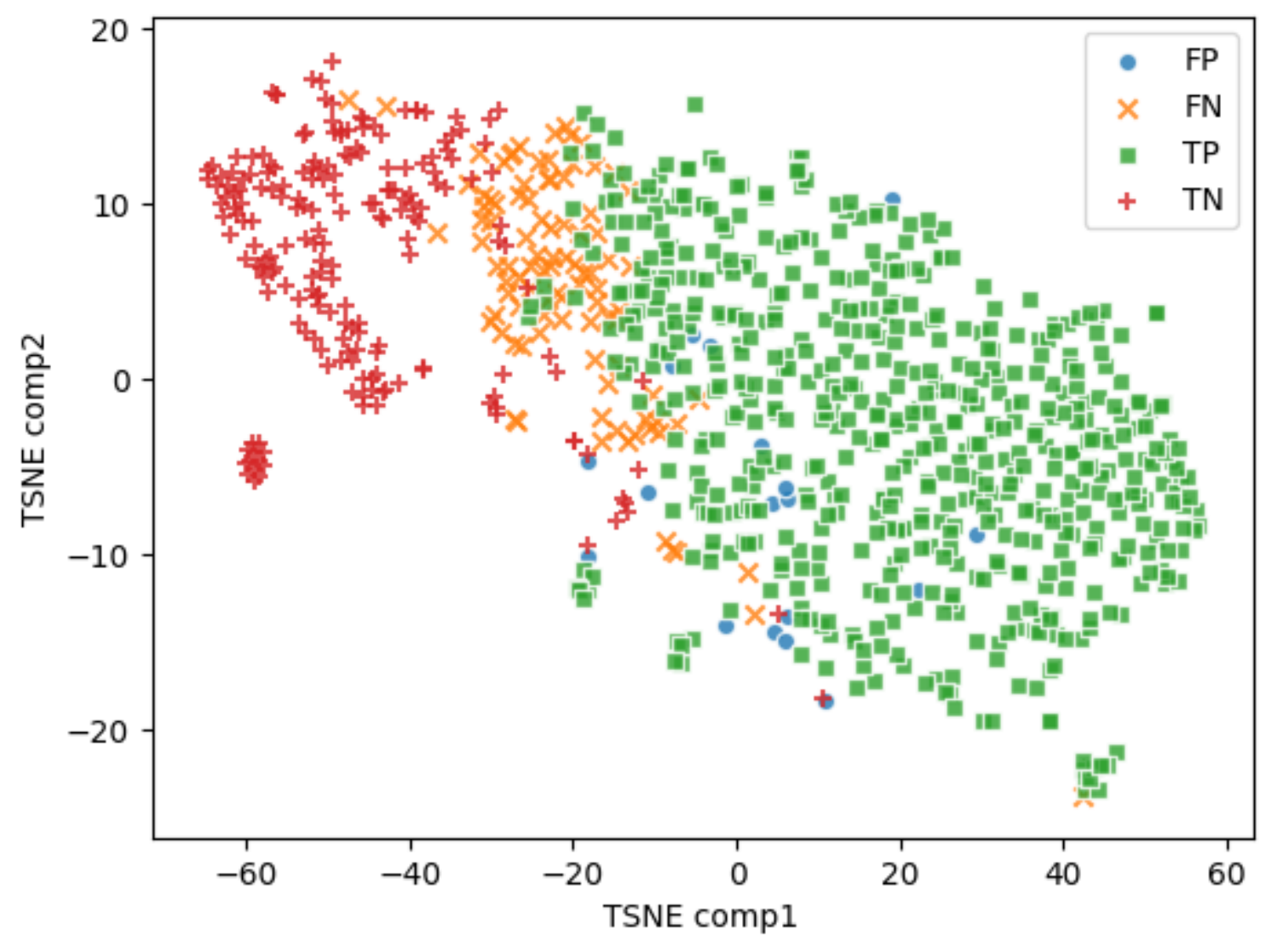} \\ 
\end{tabular}
    \caption{Visualization of the 32 features before the dense layer, reduced to two dimensions using T-SNE, for the two TOD files of the Figure~\ref{fig:cmvis}. The markers correspond to False Positives (FP, blue circle), False Negatives (FN, orange cross), True Positives (TP, green square), and True Negatives (TN, red plus). The left panel shows that the two classes of detectors are completely separated, forming two clusters. On the other hand, the right panel shows the absence of well-defined clusters, explaining the rate of false positives and negatives. }
    \label{fig:tsnevis}
\end{figure*}

To understand how the model is separating the classes, we made visualizations of the high level features extracted by the model. We start by feeding the model with the test data and exporting the features generated at the global max-pooling layer. Then, the resulting vector of length 32 is reduced to two dimensions using t-Distributed Stochastic Neighbor Embedding (t-SNE,~\cite{Maaten2008tsne}). This helps us visualize how the model is grouping the data. The result is shown in Figure~\ref{fig:tsnevis}, for the same TOD files considered in Figure~\ref{fig:cmvis}. In the first case, the two classes are well separated in the two-dimensional space forming two distinctive clusters, while in the case with lower PWV, the false positives and false negatives lie between the two main classes, forming a single cluster. These examples show us the confusion induced by the noise in the labels and the ability to separate both populations in presence of different atmospheric conditions.

\subsection{Cross-band, cross-season tests}
The trained model is tested on time-streams corresponding to other frequency bands and seasons. This is done to check whether our model is learning a more general set of features that can be transferred without retraining the classifier for those specific datasets. A summary of the results is shown in Tables~\ref{tab:crossclass_summary} and~\ref{tab:crossclass_summary2}. These results show that our deep neural network might be finding conservatives quantities across the data, which translates to generalization capabilities that transform our model into a good candidate for fast classification, avoiding tedious parameter tuning for each band and season. In the cross-band test, there is a slight decrease in performance, especially for 280 GHz for which the data selection is known to be noisier. For those cases, further improvements can be done by training a model with multi-frequency (multi-season) data or by fine-tuning the existing model for a specific data set. However, we still need to evaluate these predictions in map-space to assess for systematics and noise levels.

\begin{table}
    \caption{Classification results for cross-band test. The columns show the rate and number of True Positives (TP), True Negatives (TN), False Positives (FP) and False Negatives (FN).}
    \label{tab:crossclass_summary}
    \begin{tabular}{lcccc}
    \hline
    Frequency (Season) & TP & TN & FP & FN \\
    \hline
    220 GHz (2008) & 98.7\% & 97.0\% & 3.0\% & 1.3\%  \\
                   & (183,298) & (38,873) & (1,191) & (2,400) \\
                   & & & & \\
                   & Precision & 99.4\% & & \\
                   & Recall & 98.7\% & & \\
                   & F1-Score & 99.0\% & & \\
    \hline
    280 GHz (2008) & 94.5\% & 97.5\% & 2.5\% & 5.5\%  \\
                   & (24,435) & (30,028) & (761) & (1,420) \\
                   & & & & \\
                   & Precision & 97.5\% & & \\
                   & Recall & 94.5\% & & \\
                   & F1-Score & 95.7\% & & \\
    \hline
    \end{tabular}
\end{table}

\begin{table}
    \caption{Classification results for cross-season test. The columns show the rate and number of True Positives (TP), True Negatives (TN), False Positives (FP) and False Negatives (FN).}
    \label{tab:crossclass_summary2}
    \begin{tabular}{lcccc}
    \hline
    Frequency (Season) & TP & TN & FP & FN \\
    \hline
    148 GHz (2009) & 99.4\% & 99.7\% & 0.3\% & 0.6\%  \\
                   & (370,286) & (216,444) & (712) & (2,365) \\
                   & & & & \\
                   & Precision & 99.8\% & & \\
                   & Recall & 99.4\% & & \\
                   & F1-Score & 99.6\% & & \\
    \hline
    148 GHz (2010) & 98.7\% & 99.1\% & 0.9\% & 1.3\%  \\
                   & (390,231) & (214,062) & (1,990) & (5,078) \\
                   & & & & \\
                   & Precision & 99.5\% & & \\
                   & Recall & 98.7\% & & \\
                   & F1-Score & 99.1\% & & \\
    \hline
    \end{tabular}
\end{table}

\subsection{Training and prediction time}
Table~\ref{tab:times} summarizes the relevant times. Our baseline is the average time per TOD file for the expert software, which is designed to run in CPU. For this implementation, the average classification time is around 90 seconds per TOD file. The deep learning model takes advantage of the GPU tensor computation capabilities, boosting our prediction times up to $10\times$ per TOD file, using 3 NVIDIA GTX1080 GPUs. The training and prediction can be done in parallel at data level.

\begin{table}
    \caption{Training and prediction times.}
    \label{tab:times}
    \begin{tabular}{lr}
    \hline
    Process & Time \\
    \hline
    Data read and preprocessing & $\sim$ 12\, \text{s/TOD-file} \\
    Training ($\sim$ \text{38k examples}) & $\sim$ 12\, \text{hrs} \\
    Testing & $\sim$ 0.008\, \text{s/detector} \\
    Baseline & $\sim$ 90\, \text{s/TOD-file}\\
    \hline
    \end{tabular}
\end{table}

\section{Discussion and conclusions}\label{sec:discussion}
We have developed a deep learning model to classify the detectors of the Atacama Cosmology Telescope. The model is trained in a supervised fashion, relying on expert knowledge generated with the methodology explained in Section~\ref{sec:cuts}. In contrast to the expert method, our architecture maps from the raw data to the labels without custom-designed or tuned features. Instead, the proposed model designs its features during the training phase. Additionally, our network uses only $\sim 3$ minutes of data, reducing the prediction time by a factor of $\sim10$ for an array of 1,000 detectors; this implies that the proposed model has the potential to be used as a real-time classifier, enabling fast detector performance assessment and failure detection. 

There are two main drawbacks in the proposed solution: supervision and interpretation. Supervised learning techniques require labels previously generated that are assumed to be correct. This fact could not be true since the current methodology to generate the classes is subject to arbitrariness in the definition of the limits applied to the statistical parameters determined by the expert software. In other words, our ground-truth is not tied to a formal definition of what is a good detector. Then, the success of the model training and prediction is biased in the sense that is reproducing what the expert thinks is optimal. However, we still need to evaluate our data selection in map-space in order to quantify the impact of our findings. In terms of interpretation, our model is less transparent than the expert method or, in other words, it is not clear how the decisions are made inside the neural network. Up to now, there is no a reliable method to interpret a deep learning model. Although the error prediction is as low as 1\%, we cannot give a definite explanation of it. Here are some intuitions on why it works: the input includes the common mode as a second channel, introducing some sense of similarity; the residual blocks help to propagate through the network a measure of how dissimilar the input channels are. These are open questions that need to be studied in detail to get some physical clues that guide us to reliable interpretations, and to establish some criteria for the future design of new machine learning models.

\section*{Acknowledgements}
We thank the ACT Collaboration for the access to the MBAC data and software. F. Rojas acknowledge CONICYT for his doctoral scholarship Doctorado Nacional 2016-23190411. L. Maurin acknowledges CONICYT for a postdoctoral grant FONDECYT 3170846. Also, we acknowledge the support from CONICYT-Chile, through the FONDECYT Regular project number 1180054.




\bibliographystyle{mnras}
\bibliography{biblio} 







\bsp	
\label{lastpage}
\end{document}